\documentclass[aps,prl,twocolumn,showpacs]{revtex4}
\usepackage{graphicx}   %Include figure files
\usepackage{epsfig}
\usepackage{color}
\usepackage{amsmath,amsthm}
\usepackage{dcolumn}    %Align table columns on decimal point
\usepackage{bm}         %bold math

\newcommand{\vct}[1]{\mathbf{#1}}

\begin{document}
%\preprint{(Submitted to Phys. Rev. Lett.)}

\title{Dynamical Clustering of Counterions on Flexible Polyelectrolytes}

\author{Tak Shing Lo$^1$, Boris Khusid$^3$ and Joel Koplik$^{1,2}$}

\affiliation{Benjamin Levich Institute$^1$ and Department of Physics$^2$,
City College of the City University of New York, New York, NY 10031 \\
$^3$Department of Chemical Engineering, New Jersey Institute of 
Technology, University Heights, Newark, NJ 07102}

\date{\today}

\begin{abstract}
Molecular dynamics simulations are used to study the local dynamics of 
counterion-charged polymer association at charge densities above 
and below the counterion condensation threshold. Surprisingly, the 
counterions form weakly-interacting clusters which exhibit short range 
orientational order, and which decay slowly due to migration of ions 
across the diffuse double layer. The cluster dynamics are insensitive to 
an applied electric field, and qualitatively agree with the available 
experimental data. The results are consistent with predictions of the 
classical theory only over much longer time scales.
\end{abstract}


\maketitle

Many biochemical reactions involve local interactions between small 
molecules (ligands) and biological macromolecules such as DNA, which are 
polymers consisting of repeating ionizable groups known as 
polyelectrolytes \cite{Oosawa71}. When dissolved in a polar solvent such 
as water, polyelectrolytes ionize and counterions dissociate from the 
polymer, leaving an oppositely charged polymer backbone. The highly 
ionized polyelectrolyte in solution attracts small mobile counterions of 
opposite charge, partially screening the backbone charge.
The approach of the small molecules to a polyelectrolyte is governed to a 
large degree by electrostatic interactions, which in turn depend on the 
local molecular environment, and in particular is sensitive to the 
instantaneous distribution of counterions around the polyelectrolyte. 
Thus, knowledge of the counterion distribution at atomic resolution is 
crucial for many aspects of DNA dynamics such as molecular assembly and 
inter-cellular transport.

%Polyelectrolytes are molecules consisting of repeating ionizable groups 
%which dissociate in water or other polar solvents, leaving a charged 
%polymer backbone. A highly ionized polyelectrolyte in solution attracts 
%small mobile counterions of opposite charge, partially screening the 
%backbone charge. It has been long recognized that the mediation of the 
%Coulomb interactions of charged polymers by these associated counterions 
%plays a crucial role in numerous materials and phenomena, ranging from 
%detergents and cosmetics to proteins and inter-cellular transport 
%\cite{Oosawa71}.

The difficulty in studying polyelectrolyte solutions resides in the 
delicate interplay of the long range electrostatic interactions, short 
range intermolecular interactions, and thermal energy, which are 
comparable to each other in magnitude. Furthermore, typical 
polyelectrolytes are highly charged, which precludes straightforward 
application of the usual Debye-H\"uckel theory \cite{Hunter01} of 
electrolytic solutions.
A key advance in understanding the equilibrium state of counterions 
associated with a charged polymer was the condensation theory of Manning 
\cite{Manning69}, which treated the polyelectrolyte molecule as an 
infinite rigid rod of uniform charge density $-z_p e/b$, as an 
approximation to discrete groups of equal charge $-z_p e$ separated by a 
distance $b$, and considered independent dissolved counterions of charge 
$z_c e$ placed in a uniform bulk solvent of dielectric constant 
$\epsilon_r$ about the molecule. Considering the two-dimensional partition 
function in the plane normal to the polyelectrolyte axis, Manning found 
that the free counterion configuration is unstable for sufficiently strong 
electrostatic interaction, when the dimensionless ``Manning parameter'' 
$\xi\!\equiv\!z_p z_c l_B/b\!>\!1$, where $l_B\!=\!e^2/(\epsilon_rk_BT)$ 
is the ``Bjerrum length'' at which thermal energy $k_BT$ equals 
electrostatic potential energy. He then hypothesized that counterions 
would condense {\em uniformly} on the polyelectrolyte backbone so as to 
reduce $\xi$ to unity, while the remaining (dilute) unbound ions in 
solution could be treated in the Debye-H\"uckel approximation. Since the 
seminal work of Manning, several analytical and numerical investigations 
\cite{Prabhu05} have relaxed various assumptions in the original model. 
While the qualitative aspects of Manning's predictions survive, the 
phase transition at $\xi$=1 is present only for an infinite rod.

In contrast to the average properties of the equilibrium state of 
condensed counterions, little is known about the local dynamics of 
counterion-charged polymer association, which trigger and control 
ion-exchange reactions and affect the binding affinity of polyelectrolytes 
by shielding local electrostatic forces. In particular, most previous 
theoretical work is of a mean-field nature, and certainly cannot address 
any issues relating to temporal or spatial fluctuations in the 
distribution of the counterions along a polyelectrolyte.
The electrophoresis of polyelectrolytes raises the further issue of the 
response of the charge distribution to an applied electric field; previous 
treatments assume the counterion distribution is unchanged from its 
equilibrium state, but at least at high field strengths some modification 
will occur.

In order to understand the fine scale behavior in the distribution of the 
counterions in the vicinity of a polyion, we employ molecular dynamics 
(MD) simulations \cite{AllenTildesley87} to investigate, for the first 
time, the dynamics of counterion condensation on a mobile and flexible 
polyelectrolyte in a solution. Previous numerical studies using MD, or 
Langevin dynamics \cite{LiuMuthu0203} or Monte-Carlo methods 
\cite{NajiNetz06} have focused instead on the average behavior of the 
charge distribution or on polymer dynamics issues. Our computational 
formulation is similar to that of Chang and Yethiraj \cite{ChangYethiraj} 
and is based on a bead-spring model of a flexible polymer chain 
\cite{GrestKremer86} of Lennard-Jones monomers bound by FENE interactions, 
suspended in {\em explicit} solvent molecules, some of which are charged 
to represent counterions and ions from added salt. The polyelectrolyte 
chain contains $N$=50 monomers carrying a total charge $-Z_pe$ distributed 
uniformly along the chain. For computational tractability, the solvent 
molecules are modeled as Lennard-Jones (LJ) spheres, with the 
polarizability of water accounted for through its dielectric constant 
$\epsilon_r\!=\!78$ in the Coulomb interaction between charges. The 
polyion monomers and dissolved ions are likewise treated as LJ spheres, 
with an appropriate charge at the center, and for simplicity all are 
assigned the same mass $m$ and core size $\sigma$. All dissolved ions are 
monovalent and the counterions from the salt and the polyelectrolyte are 
assumed to be identical. Accordingly, the system includes $N_-\!=\!100$ 
coions of charge $-e$ and $N_+\!=\!(N_-+Z_p)$ counterions of charge $+e$ 
so as to maintain electro-neutrality.

The simulations involve a total of 32,000 atomic units (monomers, co- and 
counter-ions and solvent) in a periodic simulation cube of side 
$L\!=\!34.2\sigma$, giving a total number density 
$\rho\!=\!0.8\sigma^{-3}$. Simulations for one case with a larger box 
size $1.65L$ gives very similar results.  From the density of water at 
room temperature, assuming for simplicity a cubic packing of spheres, one 
estimates $\sigma\!\approx\!0.38$nm. The temperature in the simulations is 
maintained at $T\!=\!1.0\varepsilon/k_B=300K$ using a Nos\'e-Hoover 
thermostat ($\varepsilon$ is the depth of the LJ potential) and gives a 
Bjerrum length $l_B\!\approx\!1.85\sigma$. Given $N_-$ above, the salt 
concentration is approximately 0.073M, and the Debye length is 
$l_D\!\equiv\!(4\pi l_B\sum z_i^2\rho_i)^{-1/2}\!\approx\!2.7\sigma$, 
where $z_i$ and $\rho_i$ are the valence and the number density of the 
simple ion species $i$.
These parameters correspond to the typical experimental situation 
$\sigma\!<\!l_B\!<\!l_D\!<\!R_g$, where $R_g$ is the radius of gyration of 
the polyelectrolyte (found to be 5-10$\sigma$). To simulate the effects of 
external electric fields, a force which is proportional to the charge is 
added to every charged particle in the system. The natural MD unit of 
electric field strength $E$ is $\varepsilon/e\sigma\!\approx\!67$kV/mm, 
and here we compare $E\!=\!0$ and 1.0. We vary the polyelectrolyte charge 
$Z_p$, such that the Manning parameter $\xi$ for our system (with 
$b\approx\sigma$, $z_c=1$ and $z_p=Z_p/N$) spans a range above and below 
the critical value 1. Lastly, the characteristic time scale in the 
simulations is $\tau\!=\!\sigma(m/\varepsilon)^{1/2}$, about 1 ps here. 
Typical simulations equilibrate for 1000$\tau$ and data is accumulated 
over a 3000$\tau$ interval.

\begin{figure}[htb]
\includegraphics[width=0.5\linewidth,angle=-90]{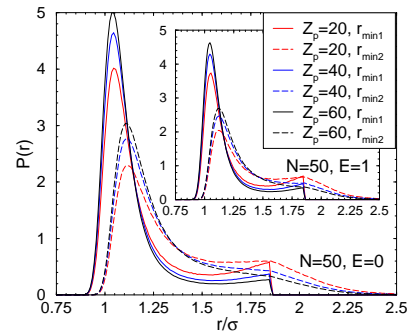} 
\caption{Probability distribution of the nearest ($r_{\rm min1}$) and the 
next nearest ($r_{\rm min2}$) distances of a bound counterion from a 
$N$=50 polyelectrolyte of charge $Z_p$=20, 40 and 60. The inset shows the 
distributions when a unit electric field is applied.}
\label{rminN50}
\end{figure}

To quantify the spatial distribution of the charge near the 
polyelectrolyte, we form a tube of radius $r$ around the polymer chain 
backbone by uniting all the spherical regions of radius $r$ centered at 
every monomer in the chain, and count the number of ions inside the 
tube. We consider a counterion to be ``bound'' to the polyelectrolyte 
molecule whenever it is within a tube of radius $l_B$; the Bjerrum length 
is the relevant length scale because within this distance the 
electrostatic interaction dominates Brownian motion and could bind the 
charges together. From the time-averaged probability distributions of the 
two nearest neighbor distances of a bound counterion to any 
polyelectrolyte monomer, shown in Fig.~\ref{rminN50}, we see that the 
bound counterions prefer to sit between two adjacent monomers forming a 
triangle. Indeed, the typical separations are very close to each other, as 
well as to the mean monomer separation ($\sim1\sigma$), and correspond to 
the three particles lying at the vertices of an equilateral triangle 
subject to modest thermal fluctuations.
Note that the distributions sharpen as the polyelectrolyte charge 
increases, and that application of an electric field of even unit strength 
has little effect.

\begin{figure}[htb]
\includegraphics[width=0.35\linewidth,angle=-90]{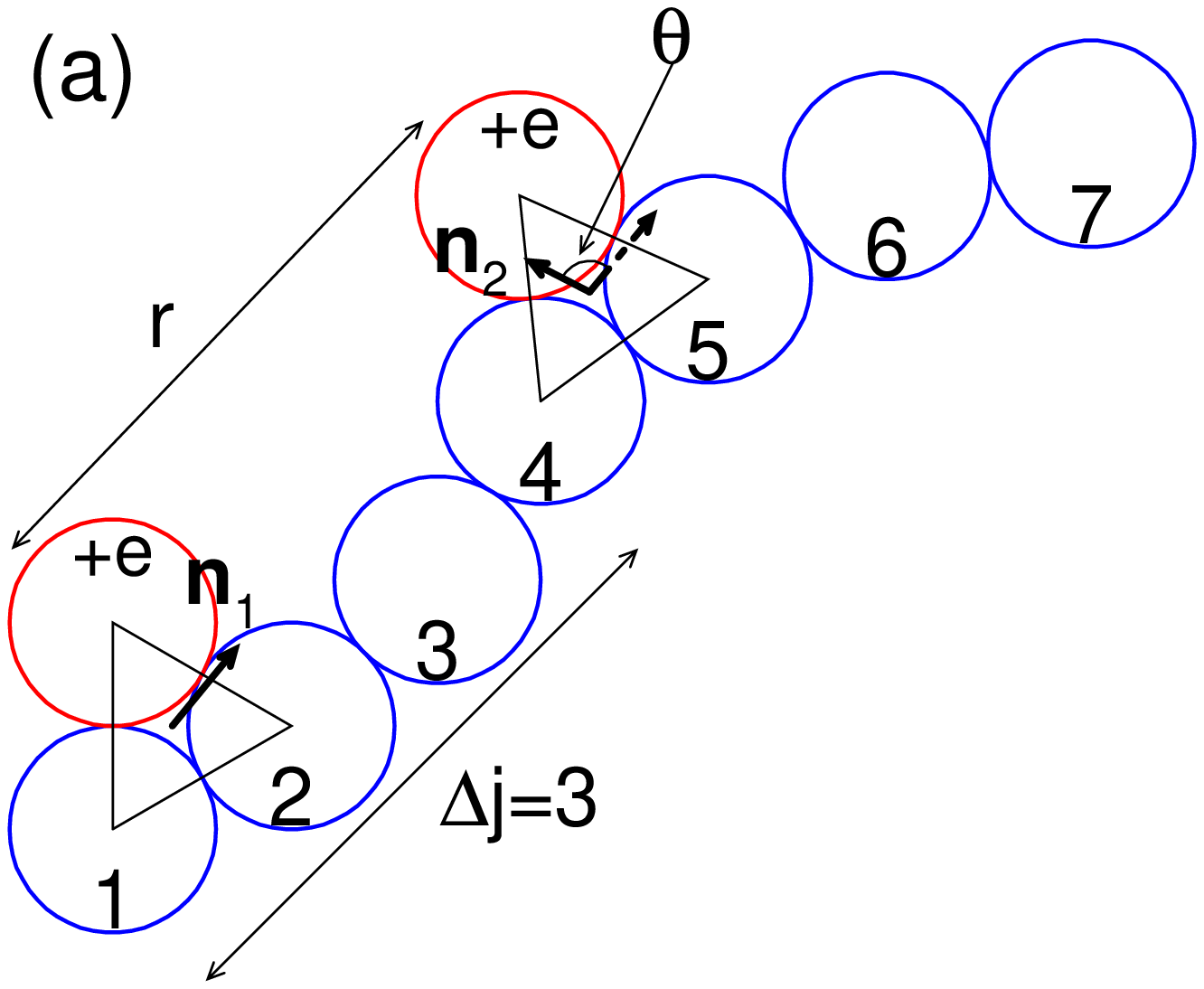}
\includegraphics[width=0.4385\linewidth,angle=-90]{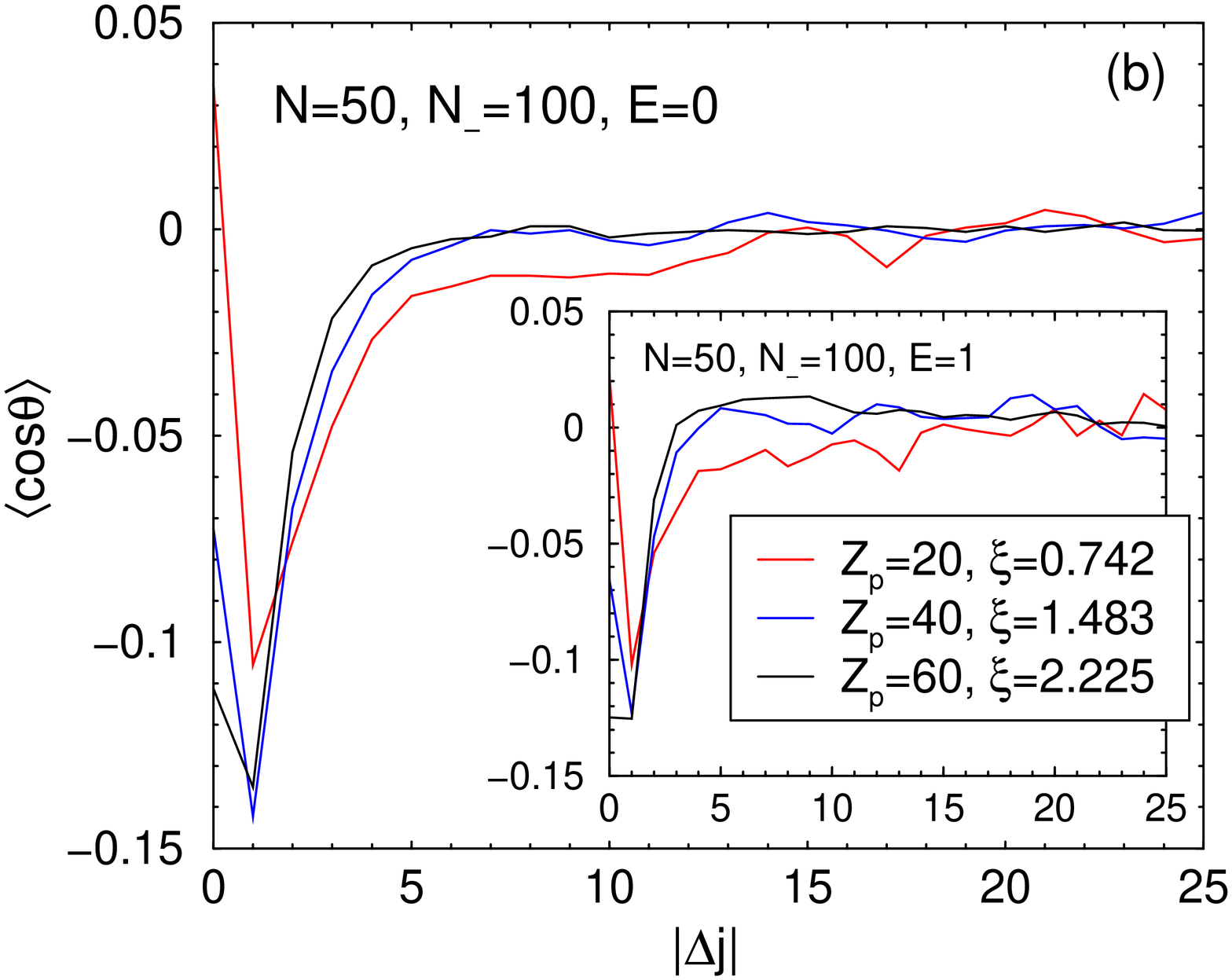} 
\caption{(a) Geometry defining the relative orientation between two 
triplets: $\theta$ is the angle between the normals $\vct{n}_1$ and 
$\vct{n}_2$ to the planes formed by two triplets. (b) Correlation in the 
relative orientations between triplets as a function of the separation 
along the molecule, with and without electric field.}
\label{CosThe}
\end{figure}

We can thus associate each bound counterion to its two nearest monomers in 
the polymer chain, which we term a monomer-ion triplet. Nearby triplets 
are strongly correlated at short length scales because the positively 
charged counterions tend to avoid each other. To characterize these 
orientational correlations, we consider the angle $\theta$ between the 
normals to the planes formed by any two triplets; see 
Fig.~\ref{CosThe}(a), and compute the time-averaged value of $\cos\theta$ 
as a function of the separation along the molecule, as defined by the 
difference $\Delta j$ in the label $j$ ($j$=1, 2,$\cdots$, $N$), of the 
first monomer in the triplet. As seen in Fig.~\ref{CosThe}(b), neighboring 
triplets are strongly anti-correlated out to a distance of about 5 
monomers, while bound counterions separated by larger separations are 
independent. The correlations are weakly dependent on $Z_p$, and not 
significantly affected by an electric field.

\begin{figure}[htb]
\includegraphics[width=0.455\linewidth,angle=-90]{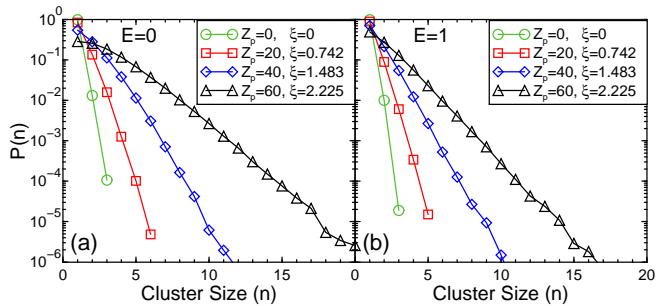}
\caption{Probability of finding a monomer-ion cluster of size $n$ for (a) 
$E$=0 and (b) $E$=1.}
\label{NclutDist}
\end{figure}

The counterion-monomer triplets exhibit dynamical clustering along the 
polyion. We define a cluster of size $n$ to consist of $n$ adjacent 
triplets along the chain backbone with no intervening vacancies, and plot 
the probability distributions $P(n)$ for clusters of size $n$ for 
different polyion charge $Z_p$ in Fig.~\ref{NclutDist}. Provided $n$ is 
not too small, we find an exponential distribution $P(n)\!\sim\!-n$, both 
with and without an electric field. Note that the range of charges 
$0\!\leq\!Z_p\!\leq\!60$, corresponds to $0\!\leq\!\xi\!\leq\!2.225$, and 
spans the Manning transition region. If we interpret this probability 
distribution as arising from a Boltzmann factor, $\ln P(n)\sim-E_n/k_BT$, 
where $E_n$ is the energy of a cluster of size $n$, then the data 
indicates that $E_n$ is approximately proportional to $n$. In consequence 
$E_{n1+n2}\!\approx\!E_{n1}$+$E_{n2}$, or in another words, the energy of 
a cluster beyond a certain (small) size depends only on its size, 
independent of its environment. Thus, clusters do not interact with each 
other and are independent. Remarkably, the cluster energies $E_n$ are not 
significantly affected by the applied field, as can be seen from the weak 
dependence of the slopes of the linear regions of the $P(n)$ curves for 
$E=0$ and $E=1$.

The connection between independence of the clusters and the exponential 
form of $P(n)$ can be seen in another way, by reference to the random 
sequential adsorption (RSA) model \cite{Evans93} for polymer reactions. If 
ions are independently adsorbed with probability $p$ and released with 
probability $q$ per unit time at different sites, then the probability of 
forming a cluster of size $n$ is proportional to $[p/(p+q)]^n$ 
\cite{cluster_prob}. The latter expression has the same exponential decay 
with $n$ as our simulation data, although this model is too simple to 
predict the slope correctly. If the ratio $p/q$ is adjusted to give the 
mean number of counterions within the Bjerrum layer separately for each 
$Z_p$ then the slope increases systematically with $Z_p$ as in our 
simulations, but does not have the correct numerical value.

\begin{figure}[htb]
\includegraphics[width=0.44\linewidth,angle=-90]{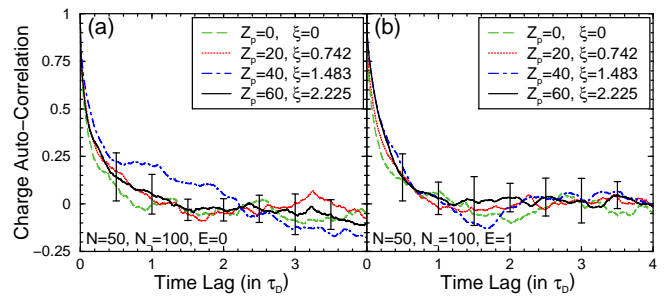}
\caption{Autocorrelation functions
$C_{QQ}(t)$=$[\langle Q(t)Q(0)\rangle\!-\!\langle Q(0)\rangle^2]/
                      [\langle Q(0)^2\rangle\!-\!\langle Q(0)\rangle^2]$ 
for the net ion charge $Q(t)$ inside $r$=$l_B$ for a polyelectrolyte of 
length $N$=50 (a) in thermodynamic equilibrium and (b) translating in the 
presence of an applied field $E$=1.0$\varepsilon/e\sigma$. Each curve is 
an average over 5 different realizations. The error bars shown correspond 
to 1 standard deviation for the case $Z_p$=60.}
\label{ChaCorrlB}
\end{figure}

Furthermore, the binding of the counterions to the chain is {\em 
transient} rather than permanent. The direct evidence comes from snapshots 
of successive configurations from the simulations, where one sees that 
bound counterions continuously move in and out of the Bjerrum layer and do 
not remain adjacent to a particular monomer indefinitely. We quantify 
these observations through the charge autocorrelation function $C_{QQ}(t)$ 
of the total ionic charge $Q(t)$ within the Bjerrum layer. As shown in 
Fig.~\ref{ChaCorrlB}, $C_{QQ}$ decays over a time close to the Debye time 
$\tau_D$ and then fluctuates about zero, indicating counterion decorrelation 
and transport through the Bjerrum layer. The relevant 
decorrelation time scale is the time $\tau_D\!=\!l_D^2/D$ for an 
isolated counterion to diffuse through a Debye length, where $D$ is the 
diffusion constant of the counterions. The latter parameter was measured 
in independent simulations without a polyion to be 
$D\!=\!0.070\sigma^2/\tau$, leading to $\tau_D\!\approx\!0.1$ns.
Note in particular that there is no significant difference between a 
neutral and charged polyelectrolyte molecule, and again an applied 
electric field has little effect. 

\begin{figure}[htb]
\includegraphics[width=0.46\linewidth,angle=-90]{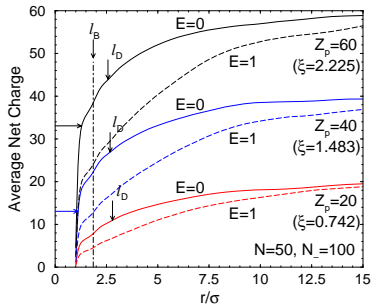} 
\caption{Charge distribution around a polyion for $Z_p$=20, 40 and 60 with 
and without an applied electric field. The horizontal arrows indicate the 
charge of condensed counterions predicted by Manning's theory in each case 
for $\xi$$>$1.}
\label{ChaDist}
\end{figure}

Aside from the fluctuations apparent in $C_{QQ}$, the spatial distribution 
of charge within the Bjerrum layer is constant on average, and shows an 
approximate agreement with Manning's predictions. Fig.~\ref{ChaDist} shows 
the total charge within a tube of radius $r$ about polyions of charge 
$Z_p=20$, 40 and 60, over a time interval of $3000\tau\!\gg\!\tau_D$ in 
the steady state. In the latter two cases, the linear charge densities of 
the polyion are above the critical value for Manning condensation. In the 
absence of an electric field, we see that the time-averaged total charge 
of the ions within the Bjerrum layer when $\xi\!>\!1$ approximately agrees 
with Manning's prediction.
The original model was imprecise about exactly where the condensed 
counterions would be located; we find that the condensed charge in the 
region $r\!\lesssim\!1.5\sigma$ is very close to the prediction. Moreover, 
the number of the coions within the Debye layer is found to be negligibly 
small as was assumed in the theory. Furthermore, the slopes of the solid 
curves for $E=0$ in Fig.~\ref{ChaDist} for the cases $Z_p$=40 and 60 above 
the transition point are almost the same for $r\!>\!l_D$. This fact is 
qualitatively consistent with Manning's treatment, where the ``unbound'' 
counterions are treated by the linear Debye-H\"uckel theory, neglecting 
the effects of the polyelectrolyte charge.
Also as indicated in Fig.~\ref{ChaDist}, the effect of the applied 
electric field is to reduce the charge in the vicinity of the 
polyelectrolyte compared to the zero-field value, in contrast to Manning's 
theory of the electrophoresis of polyelectrolytes for $\xi>1$ 
\cite{Manning81} and its generalizations
\cite{Cleland91th,VolkelNoolandi95,MohantyStellwagen99,Hoagland99}.

Experiments have just begun to probe counterion structure and dynamics in 
polyelectrolyte solutions at nanometer and nanosecond scales. Small-angle 
neutron scattering measurements of the counterion-counterion structure 
factor \cite{Prabhu04} indicate that the counterions are strongly 
correlated with the polymer chain so as to mimic the pair correlations 
typical for the polymer, indicating that counterions are localized in the 
vicinity of the polyions.
Similarly, small-angle neutron and X-ray measurements \cite{Morfin04} 
provide evidence for counterion localization below the Debye layer, and 
inelastic X-ray scattering measurements \cite{Angelini06} find phonon-like 
excitations in the counterion distribution, indicating the presence of a 
correlated structure.
Electron paramagnetic resonance spectroscopy studies \cite{Hinderberger03} 
of counterion dynamics gave direct evidence for the existence of preferred 
sites corresponding to the charged groups of the polyelectrolyte, where 
the counterions are electrostatically attached directly with no solvent 
molecule in between. Such ``site bound'' counterions presumably correspond 
to the bound counterions inside the Bjerrum layer observed in our 
simulations. We find that such counterions enter and leave the Bjerrum 
layer with a time scale around $\tau_D\approx0.1$ns, while the experiments 
find that the exchange rate constant for the the ``site-bound'' 
counterions is significantly larger than $10^9$s$^{-1}$, implying a 
lifetime shorter than 1ns, consistent with our result.
Lastly, recent experimental findings related to polyelectrolyte 
electrophoretic mobility \cite{PopovHogland04} argue that an ion pairing 
mechanism (Bjerrum association \cite{Levin02}) resembling triplet 
formation may play a role in the process of counterion condensation.

In summary, our simulations provide insight into the previously unexplored 
local dynamics of counterion condensation on a mobile and flexible 
polyelectrolyte, both above and below the condensation transition. We find 
that strong long-ranged Coulomb interactions cause counterions to form 
weakly-interacting transient clusters around the polyelectrolyte, which 
exhibit short range orientational order. The counterion clustering is 
robust to changes in the degree of ionization as well as to the 
application of an external electric field, and the cluster decay results 
from ion diffusion in and out of the charged double layer about the 
polyelectrolyte. Both the presence of clusters and their finite lifetime 
is qualitatively consistent with experimental data. The classical 
counterion condensation theory is {\em only} consistent with the local 
charge distribution at much longer time scales. Our results provide a 
framework for further theoretical and experimental studies of the dynamics 
of counterion-charged polymer association at nanometer and nanosecond 
scales, information essential for understanding the electrophoretic 
behavior of these molecules, as well as the interaction and binding of 
ligands to biological polyelectrolytes.

This work was supported in part by grants NSF CTS-0307099 and NSF/Sandia 
NIRT/NER-0330703.


\begin{thebibliography}{99}
\bibitem{cluster_prob} If $P_1(t)$ is the probability for a single site to 
be occupied at time $t$, then $P_1(t+\Delta t)$=$P_1(t)+p\Delta 
t[1-P_1(t)]-q\Delta t P_1(t)$, which for $\Delta t\to 0$ yields 
$\dot{P}_1$=$p-(p+q)P_1$. The steady state solution is $P_1$=$p/(p+q)$.
%Integrating, and taking the limit $t\to\infty$ gives $P_1\to p/(p+q)$.
If the $n$ sites are independent, the result follows.
\end{thebibliography}
\end{document}